\documentclass[twocolumn,english,american]{revtex4-2}
\usepackage[T1]{fontenc}
\usepackage[utf8]{luainputenc}
\usepackage{geometry}
\usepackage{amsmath}
\usepackage{amssymb}
\usepackage{babel}
\begin{document}
\title{Linearized analysis of dissipative Two Axis Counter Twisting (TACT)
squeezing for Metrology}
\author{Garry Goldstein$^{1}$}
\address{$^{1}$garrygoldsteinwinnipeg@gmail.com}
\begin{abstract}
In this work we analyze two axis twisting in the presence of depolarizing
channel dissipation. We find that spin squeezing is only possible
if the dissipation is parametrically weaker then the squeezing coupling.
Squeezing may be used for meteorologically useful decrease of spin
noise but only in the case where the squeezing occurs before measurement,
in the case one squeezes as one measures one also squeezes the signal
thereby making spin squeezing ineffective for metrological gain. The
key mathematical advance made in this work is the observation that
TACT in the presence of depolarizing noise is equivalent to TACT with
reduced polarization and no noise. We find an exponential gain in
signal to noise with the exponent proportional to the ratio between
the squeezing strength and the depolarization rate.
\end{abstract}
\maketitle

\section{Introduction}\label{sec:Introduction}

The main resource needed for quantum computation is entanglement \citep{Nielsen_2011}.
Squeezed spin states are a resource for quantum entangled states \citep{Kitagawa_1993,Wineland_1992,Wineland_1994,Wang_2017,Polzik_2008,Guehne_2009,Bigelow_2001,Sorensen_2001,Borregaard_2017}.
One Axis Twisting (OAT) \citep{Kitagawa_1993,Li_2008,Li_2009,Ma_2011}
and Two Axis Counter Twisting (TACT) states \citep{Borregaard_2017,Wang_2017,Kitagawa_1993}
are a resource of entanglement particularly useful for precision metrology
at the sub shot noise level \citep{Wineland_1992,Wineland_1994,Li_2009,Li_2008}.
The quantum uncertainty of measurement is given by $\xi/\sqrt{N}$
where $\xi$ is the squeezing parameter and $1/\sqrt{N}$ is the shot
noise limit \citep{Cronin_2009}. Here $N$ is the total number of
spins \citep{Kitagawa_1993,Wineland_1992,Wineland_1994}. For OAT
squeezing the Hamiltonian is given by a non-linear single axis interaction
$H\sim S_{z}^{2}$ (here $S$ is the total spin for the $N$ constituent
spin one halves) \citep{Kitagawa_1993}, while for TACT the Hamiltonian
is a sum of two non-linear spin interactions $H\sim S_{x}^{2}-S_{y}^{2}$
\citep{Kitagawa_1993,Wang_2017,Borregaard_2017}. Both OAT and TACT
Hamiltonians may be obtained experimentally. OAT may be obtained from
the Dicke model with quantum non-demolition measurement, transfer
of squeezing from light to spin ensembles and with the use of atom-atom
interactions in Bose Einstein Condensates (BEC's) \citep{Ma_2011}.
OAT squeezing has been experimentally demonstrated however TACT squeezing
has not though several theoretical methods to obtain TACT Hamiltonians
have been proposed \citep{Borregaard_2017,Wang_2017}. The usefulness
of TACT squeezing for metrological gain in the absence of important
decoherence sources has been well demonstrated theoretically \citep{Kitagawa_1993,Borregaard_2017,Wang_2017}.In
this work we extend this analysis by studying TACT squeezing in the
presence of decoherence in the form of a depolarizing channel. We
show analytically that TACT can lead to meteorologically useful squeezing
in the presence of depolarizing noise assuming the squeezing Hamiltonian
is parametrically greater then the depolarization rate (see Eq. (\ref{eq:Condition})).
Furthermore we show that squeezing while acquiring signal is an inefficient
measurement protocol as the squeezing Hamiltonian also squeezes the
signal, however acquiring signal after measurement even in the presence
of decoherence is a viable option for sub shot noise metrology. The
key mathematical step in this work is to show that for a depolarizing
channel studying TACT spin squeezing is equivalent to studying TACT
spin squeezing without decoherence but for reduced polarization. We
find an overall metrological improvement of $\sim\frac{\exp\left(\alpha e^{-1}\right)}{\alpha}$
where $\alpha=\frac{JNP}{4\Gamma}$. Here $J$ is the squeezing rate,
$N$ is the number of spins, $P$ is the initial polarization and
$\Gamma$ is the depolarization rate.

\section{Squeezing}\label{sec:Squeezing}

We consider a spin ensemble with $N$ spins with the initial density
matrix given by 
\begin{equation}
\rho=\otimes_{i=1}^{N}\left(\frac{\mathbb{I}}{2}+P\sigma_{z}^{i}\right)\label{eq:Density}
\end{equation}
Here $\sigma^{i}$ are the Pauli matrices. This means that the spins
are uniformly polarized along the z-axis. We consider the following
Linblad evolution for the system:
\begin{align}
\frac{\partial\rho}{\partial t} & \equiv\mathcal{L}\left(\rho\right)=\mathcal{L}_{1}\left(\rho\right)+\mathcal{L}_{2}\left(\rho\right)\nonumber \\
\mathcal{L}_{1}\left(\rho\right) & =-i\left[H_{Squ},\rho\right]\nonumber \\
\mathcal{L}_{2}\left(\rho\right) & =\Gamma\sum_{i}\left[\sigma_{x}^{i}\rho\sigma_{x}^{i}+\sigma_{z}^{i}\rho\sigma_{z}^{i}+\sigma_{y}^{i}\rho\sigma_{y}^{i}\right]-3\Gamma\rho\nonumber \\
H_{Squ} & =J\sum_{i,j}\sigma_{x}^{i}\sigma_{x}^{j}-J\sum_{i,j}\sigma_{y}^{i}\sigma_{y}^{j}\label{eq:Linbladian-1}
\end{align}
It is given by a depolarizing channel with rate $\Gamma$ and TACT
evolution with squeezing strength $J$. We now see that after time
$T$:
\begin{align}
\rho\left(T\right) & =\exp\left[T\left(\mathcal{L}_{1}+\mathcal{L}_{2}\right)\right]\rho\nonumber \\
 & =\exp\left[T\mathcal{L}_{1}\right]\exp\left[T\mathcal{L}_{2}\right]\rho\label{eq:Factorization}
\end{align}
 Where we have used that:
\begin{equation}
\left[\mathcal{L}_{1},\mathcal{L}_{2}\right]\cong0+O\left(\frac{1}{N}\right)\label{eq:Commutator_zero}
\end{equation}
Now we see that 
\begin{equation}
\exp\left[T\mathcal{L}_{2}\right]\rho=\otimes_{i=1}^{N}\left(\frac{\mathbb{I}}{2}+P\exp\left[-4\Gamma T\right]\sigma_{z}^{i}\right)\label{eq:Loss_phase}
\end{equation}
As such we have reduced the squeezing under decoherence problem to
a pure squeezing problem with finite initial polarization. To make
further analytical progress we consider the linearized Holstein Primakov
representation of spin: 
\begin{align}
S_{+} & =\sqrt{N-n_{b}}b\cong\sqrt{N\mathcal{P}}b\nonumber \\
S_{-} & =b^{\dagger}\sqrt{N-n_{b}}\cong\sqrt{N\mathcal{P}}b^{\dagger}\nonumber \\
S_{z} & =N-b^{\dagger}b\label{eq:Linearized_holstein_primakoff}
\end{align}
With 
\begin{equation}
\mathcal{P}=P\exp\left[-4\Gamma T\right]\label{eq:Polarization}
\end{equation}
Now we write: 
\begin{equation}
H_{Squ}=J\sum_{i,j}\sigma_{x}^{i}\sigma_{x}^{j}-J\sum_{i,j}\sigma_{y}^{i}\sigma_{y}^{j}=\frac{1}{2}JN\mathcal{P}\left[b^{\dagger}b^{\dagger}+bb\right]\label{eq:Squeezing}
\end{equation}
Furthermore: 
\begin{align}
\frac{\partial b}{\partial t} & =i\left[H_{Squ},b\right]=iJN\mathcal{P}b^{\dagger}\nonumber \\
\frac{\partial b^{\dagger}}{\partial t} & =i\left[H_{Squ},b\right]=-iJN\mathcal{P}b\label{eq:EOM}
\end{align}
or 
\begin{equation}
\frac{\partial}{\partial t}\left(\begin{array}{c}
b\\
b^{\dagger}
\end{array}\right)=JN\mathcal{P}\left(\begin{array}{cc}
0 & i\\
-i & 0
\end{array}\right)\left(\begin{array}{c}
b\\
b^{\dagger}
\end{array}\right)\label{eq:Squeeze}
\end{equation}
There are two eigenvalues $\lambda_{\pm}=\pm JN\mathcal{P}$ with
eigenvectors: 
\begin{equation}
v_{+}=\frac{1}{\sqrt{2}}\left(b-ib^{\dagger}\right),\:v_{-}=\frac{1}{\sqrt{2}}\left(b+ib^{\dagger}\right)\label{eq:Solutions}
\end{equation}
As such we have that 
\begin{align}
b\left(t\right) & =b\mathrm{cosh}\left(JN\mathcal{P}t\right)-ib^{\dagger}\sinh\left(JN\mathcal{P}t\right)\nonumber \\
b^{\dagger}\left(t\right) & =ib^{\dagger}\cosh\left(JN\mathcal{P}t\right)-b\sinh\left(JN\mathcal{P}t\right)\label{eq:Solution}
\end{align}
As such after time $T$ we have that: 
\begin{equation}
\xi_{min}^{2}=\frac{\exp\left(-JNP\exp\left(-4\Gamma T\right)T\right)}{P\exp\left(-4\Gamma T\right)}\label{eq:Final_answer}
\end{equation}
Writing $\Theta=4\Gamma T$ and $\alpha\Theta=JNPT$ we have that:
\begin{equation}
\xi_{min}^{2}=P^{-1}\exp\left(-\Theta\left[\alpha\exp\left(-\Theta\right)-1\right]\right)\label{eq:Squeezing-1}
\end{equation}
As such its sufficient to optimize 
\begin{align}
 & \Theta\left[\alpha\exp\left(-\Theta\right)-1\right]\nonumber \\
 & \Rightarrow\alpha>1\label{eq:Angle}
\end{align}
As such squeezing is only possible for 
\begin{equation}
JNP>4\Gamma\label{eq:Condition}
\end{equation}
Otherwise it is not worth to spend any time squeezing at all and measure
with unsqueezed states. The squeezing in Eq. (\ref{eq:Final_answer})
can now be optimized numerically. For qualitative understanding we
now work in the strong squeezing limit, $\alpha\gg1$ which means
that $\Theta_{min}\cong1$ and 
\begin{equation}
\xi_{min}^{2}=P^{-1}\exp\left(-\left[\alpha e^{-1}-1\right]\right)\label{eq:Approximate}
\end{equation}
This shows improved squeezing when the condition in Eq. (\ref{eq:Condition})
is met.

\section{Adding signal: metrology applications}\label{sec:Adding-signal-for}

\subsection{Squeezing while measuring}\label{subsec:Squeezing-while-measuring}

Let us now add an external magnetic field. More precisely let us consider
the evolution: 
\begin{align}
\frac{\partial\rho}{\partial t} & \equiv\mathcal{L}\left(\rho\right)=\mathcal{L}_{1}\left(\rho\right)+\mathcal{L}_{2}\left(\rho\right)+\mathcal{L}_{3}\left(\rho\right)\nonumber \\
\mathcal{L}_{3}\left(\rho\right) & =i\left[B\sum_{i}\left[\sigma_{y}^{i}-\sigma_{x}^{i}\right],\rho\right]\label{eq:Field}
\end{align}
Now we see that after time $T$:
\begin{align}
\rho\left(T\right) & =\exp\left[T\left(\mathcal{L}_{1}+\mathcal{L}_{2}+\mathcal{L}_{3}\right)\right]\rho\nonumber \\
 & =\exp\left[T\mathcal{L}_{1}+\mathcal{L}_{3}\right]\exp\left[T\mathcal{L}_{2}\right]\rho\label{eq:Factorization-1}
\end{align}
 Where we have used that:
\begin{equation}
\left[\mathcal{L}_{3},\mathcal{L}_{2}\right]=0\label{eq:Commutator_zero-1}
\end{equation}
As such we see that squeezing while acquiring signal with depolarizing
channel decoherence is equivalent to squeezing while acquiring signal
in a partially polarized state. Therefore (using Eq. (\ref{eq:Linearized_holstein_primakoff})
the total Hamiltonian for the squeezing while acquiring signal is
given by: 
\begin{align}
H & =J\sum_{i,j}\sigma_{x}^{i}\sigma_{x}^{j}-J\sum_{i,j}\sigma_{y}^{i}\sigma_{y}^{j}-B\sum_{i}\left[\sigma_{y}^{i}-\sigma_{x}^{i}\right]\nonumber \\
 & =\frac{1}{2}JN\mathcal{P}\left[b^{\dagger}b^{\dagger}+bb\right]-B\sqrt{N\mathcal{P}}\left(\frac{1+i}{2}b-\frac{1-i}{2}b^{\dagger}\right)\nonumber \\
 & =\frac{1}{2}JN\mathcal{P}\left[\mathcal{B}^{\dagger}\mathcal{B}^{\dagger}+\mathcal{B}\mathcal{B}\right]+const\label{eq:Hamiltonian}
\end{align}
Where 
\begin{align}
\mathcal{B}^{\dagger} & =b^{\dagger}-\frac{B}{2J\sqrt{N\mathcal{P}}}\left(1-i\right)\nonumber \\
\mathcal{B} & =b-\frac{B}{2J\sqrt{N\mathcal{P}}}\left(1+i\right)\label{eq:Field_shifts}
\end{align}
This means that after a time $t$:
\begin{align}
v_{+}^{\mathcal{B}}\left(t\right) & =v_{+}\exp\left(JN\mathcal{P}t\right)\nonumber \\
v_{-}^{\mathcal{B}}\left(t\right) & =v_{-}\exp\left(-JN\mathcal{P}t\right)\label{eq:Solution-1}
\end{align}
Where 
\begin{equation}
v_{+}^{\mathcal{B}}=\frac{1}{\sqrt{2}}\left(\mathcal{B}-i\mathcal{B}^{\dagger}\right),\:v_{-}^{\mathcal{B}}=\frac{1}{\sqrt{2}}\left(\mathcal{B}+i\mathcal{B}^{\dagger}\right)\label{eq:Solutions-1}
\end{equation}
This means that 
\begin{align}
\left\langle v_{-}^{\mathcal{B}}\left(t\right)\right\rangle  & =-\exp\left(JN\mathcal{P}t\right)\frac{B}{J\sqrt{N\mathcal{P}}}\left(1+i\right)\nonumber \\
\left\langle v_{+}^{\mathcal{B}}\left(t\right)\right\rangle  & =0\label{eq:Zero}
\end{align}
Here we have used the initial state in Eq. (\ref{eq:Loss_phase}).
This means that 
\begin{align}
\left\langle v_{-}\left(t\right)\right\rangle  & =\left[1-\exp\left(JN\mathcal{P}t\right)\right]\frac{B}{J\sqrt{N\mathcal{P}}}\left(1+i\right)\nonumber \\
\left\langle v_{+}\left(t\right)\right\rangle  & =0\label{eq:Values}
\end{align}
Now the signal is given by: 
\begin{align}
\mathcal{SIG} & =\left\langle S_{x}+S_{y}\right\rangle =\frac{\sqrt{N\mathcal{P}}}{\left(1+i\right)}\left\langle v_{-}\left(t\right)\right\rangle \nonumber \\
 & =\frac{B}{J}\left[1-\exp\left(-JN\mathcal{P}t\right)\right]\label{eq:Signal}
\end{align}
This means that for a total measurement time $\tau$ with each measurement
taking a time of $T$ for squeezing and the signal to noise is given
by: 
\begin{equation}
\frac{1}{\sqrt{\tau}}\frac{\partial\mathcal{S}}{\partial B}=\frac{\sqrt{2}}{J\sqrt{TN}}\frac{1-\exp\left(-JNP\exp\left(-4\Gamma T\right)T\right)}{\exp\left(-\left[JNPT-1\right]\exp\left(-4\Gamma T\right)\right)}\label{eq:Signal_noise}
\end{equation}
we see that because the squeezing also squeezes the signal spin squeezing
is ineffective for metrology.

\subsection{Squeezing before measurement}\label{subsec:Squeezing-before-measurement}

Consider now squeezing for a time $T$ and then acquiring signal for
a time $t$. That is we consider the following Linbladian evolution:
\begin{align}
\rho\left(T+t\right) & =\exp\left(\left[\mathcal{L}_{2}+\mathcal{L}_{3}\right]t\right)\exp\left(\left[\mathcal{L}_{1}+\mathcal{L}_{2}\right]T\right)\rho\nonumber \\
 & =\exp\left(\mathcal{L}_{3}t\right)\exp\left(\mathcal{L}_{1}T\right)\exp\left(\mathcal{L}_{2}\left(T+t\right)\right)\rho\label{eq:Manipulations}
\end{align}
Therefore we have that acquiring signal under decoherence after squeezing
is equivalent to acquiring signal with no decoherence with a state
thats squeezed as 
\[
\xi_{min}^{2}=\frac{\exp\left(-JNP\exp\left(-4\Gamma\left(T+t\right)\right)T\right)}{P\exp\left(-4\Gamma\left(T+t\right)\right)}
\]
 In this case we have that the signal to noise satisfies: 
\begin{equation}
\frac{1}{\sqrt{\tau}}\frac{\partial\mathcal{S}}{\partial B}=\frac{t\sqrt{N}}{\sqrt{T+t}}\frac{P\exp\left(-4\Gamma\left(T+t\right)\right)}{\exp\left(-JNP\exp\left(-4\Gamma\left(T+t\right)\right)T\right)}\label{eq:Squeezing-3}
\end{equation}
This can be optimized numerically however to gain intuition we now
work in the case where $\alpha\gg1$ in which case $4\Gamma\left(T+t\right)\cong1$
and 
\begin{align}
\frac{1}{\sqrt{\tau}}\frac{\partial\mathcal{S}}{\partial B} & \cong\frac{t\sqrt{N}}{\sqrt{T+t}}\frac{Pe^{-1}}{\exp\left(-\alpha e^{-1}\left(\frac{T}{t+T}\right)\right)}\nonumber \\
 & =\frac{\sqrt{N}}{\sqrt{4\Gamma}}\frac{Pe^{-1}}{\exp\left(-\alpha e^{-1}U\right)}\left(1-U\right)\nonumber \\
\frac{1}{\sqrt{\tau}}\frac{\partial\mathcal{S}_{max}}{\partial B} & =\frac{\sqrt{N}}{\sqrt{4\Gamma}}\frac{P\alpha^{-1}}{\exp\left(-\left[\alpha e^{-1}-1\right]\right)}\label{eq:Optimum}
\end{align}
Where $U=4\Gamma T$ and $U_{max}=\frac{\alpha e^{-1}-1}{\alpha e^{-1}}$.
We notice that without squeezing $\frac{1}{\sqrt{\tau}}\frac{\partial\mathcal{S}_{max}}{\partial B}\sim P\frac{\sqrt{N}}{\sqrt{\Gamma}}$,
which means the metrological improvement is given by: $\sim\frac{\exp\left(\alpha e^{-1}\right)}{\alpha}$.

\section{Conclusion}\label{sec:Conclusion}

In this work we have studied TACT spin squeezing in the presence of
depolarizing channel noise. We have shown that in the case where the
squeezing Hamiltonian is parametrically greater then the depolarizing
decoherence rate meteorologically useful squeezing may be achieved
(see Eq. (\ref{eq:Condition})). Spin squeezing while acquiring signal
on the other hand is not a promising pathway for sub shot noise meteorology
as the signal gets squeezed as well as the noise. However pre-squeezing
followed by signal acquisition even in the presence of a depolarizing
decoherence channel is a viable method to obtain sub shot noise meteorology.
These results further support TACT as a method of spin squeezing for
meteorological gain. The overall metrological improvement scales as
$\sim\frac{\exp\left(\alpha e^{-1}\right)}{\alpha}$ where $\alpha=\frac{JNP}{4\Gamma}$. 

\appendix

\section{Argument why Eq. (\ref{eq:Commutator_zero}) is reasonable}\label{sec:Argument-why-Eq.}

We consider the case of a large number of spins. In which case we
have that for a term in the density matrix: 
\begin{equation}
\rho=\sum_{\alpha}P_{\alpha_{i}}\prod\sigma_{\alpha_{i}}^{i},\quad\alpha_{i}=\mathbb{I},x,y,z\label{eq:Density-1}
\end{equation}
Because the polarization is non-zero there is a large number $\sim N$
terms with $\alpha_{i}=x,y,z$. As such we have that 
\begin{equation}
\mathcal{L}_{2}\left(\rho\right)=-4\Gamma\sum_{\alpha}P_{\alpha_{i}}\prod\sigma_{\alpha_{i}}^{i}\left(1-\delta_{\alpha_{i},\mathbb{I}}\right)\label{eq:Decoherence}
\end{equation}
contains a large number of terms. Now we have that any terms of the
form: 
\begin{equation}
\mathcal{L}_{1x/y}^{ij}\left(\rho\right):\sum_{i}\left(1-\delta_{\alpha_{i},\mathbb{I}}\right)\rightarrow\sum_{i}\left(1-\delta_{\alpha_{i},\mathbb{I}}\right)+\left(0/+1/-1\right)\label{eq:Change}
\end{equation}
where 
\begin{equation}
\mathcal{L}_{1x/y}^{ij}\left(\rho\right)=J\sigma_{x/y}^{i}\sigma_{x/y}^{j}\label{eq:L_1}
\end{equation}
As such the rate of decoherence does not significantly change for
a typical term in Eq. (\ref{eq:Decoherence}). Furthermore we know
that for $\alpha\gg1$ we have that $4\Gamma T=1$, $JPT=\alpha/N\ll1$
so that each individual term of the form $\mathcal{L}_{1x/y}^{ij}$
need only act once during evolution (with corrections scaling as $1/N$)
so we have that: 
\begin{equation}
\left[\mathcal{L}_{1x/y}^{ij},\mathcal{L}_{2}\right]\cong0\label{eq:Zero-1}
\end{equation}
which implies Eq. (\ref{eq:Commutator_zero}).
\selectlanguage{english}%

\end{document}